\documentclass[12pt]{article}
\usepackage{amsmath,amssymb}

\usepackage{graphicx}
\usepackage{wrapfig}

\usepackage{hyperref}
\usepackage{cite}

\def\beq{\begin{eqnarray}}
\def\eeq{\end{eqnarray}}

\def\la{\langle }
\def\ra{\rangle }

\newcommand{\Tr}{\,\mathrm{Tr}\,}            

\newcommand{\be}{\begin{equation}}
\newcommand{\ee}{\end{equation}}
\newcommand{\bea}{\begin{eqnarray}}
\newcommand{\eea}{\end{eqnarray}}
\newcommand{\bg}{\begin{gather}}

\newcommand{\bseq}{\begin{subequations}}
\newcommand{\eseq}{\end{subequations}}

\def\be{\begin{eqnarray}}
\def\ee{\end{eqnarray}}
\def\lb{\label}

\begin{document}

\title{\textbf{Boundary terms of conformal anomaly }}

\vspace{2cm}
\author{ \textbf{   Sergey N. Solodukhin}} 
\date{}
\maketitle
\begin{center}
  \emph{  Laboratoire de Math\'ematiques et Physique Th\'eorique  CNRS-UMR
7350, }\\
  \emph{F\'ed\'eration Denis Poisson, Universit\'e Fran\c cois-Rabelais Tours,  }\\
  \emph{Parc de Grandmont, 37200 Tours, France}
\end{center}



\begin{abstract}
\noindent { We analyze the structure of the boundary terms in the conformal anomaly integrated  over a manifold with boundaries.
We suggest that the anomalies of type B, polynomial in the Weyl tensor, are accompanied with the respective boundary terms  of the Gibbons-Hawking type.
Their form is dictated by the requirement that they produce a  variation which  compensates  the normal derivatives of the  metric 
variation on the boundary in order  to have a well-defined variational procedure. This suggestion agrees with recent findings in four dimensions for free fields of various spin. 
We generalize this consideration to six dimensions and derive explicitly  the respective boundary terms. We point out that the integrated conformal anomaly in odd dimensions is non-vanishing due to the boundary terms. These terms are specified in three and five dimensions.
}
\end{abstract}

\vskip 1 cm
\noindent
\rule{7.7 cm}{.5 pt}\\
\noindent 
\noindent
\noindent ~~~ {\footnotesize e-mail:  Sergey.Solodukhin@lmpt.univ-tours.fr}


\newpage

\section{ Introduction}
As is well-known the variational principle for the bulk action which includes functions of the Riemann curvature 
is not well defined in the presence of boundaries. The variation of the curvature produces a normal derivative of the
metric variation on the boundary. The elimination  of this term by fixing, additionally to the metric itself on the boundary, also its normal derivative
makes the problem over constrained so that no non-trivial solution to the corresponding field equations exists. A way out was found by Gibbons and Hawking \cite{Gibbons:1976ue}
in 1977. They suggested that one has to add a boundary term which depends on the extrinsic curvature of the boundary. The role of this term is to cancel the unwanted  normal derivatives of the variation of metric. This term for the Einstein-Hilbert action, linear in the curvature,  is now known as the Gibbons-Hawking term.

For more general functions which may include  polynomials and derivatives of the curvature the appropriate boundary term was found in \cite{Barvinsky:1995dp}. 
In \cite{Barvinsky:1995dp} it was used the fact that, by adding auxiliary fields,  any function of the curvature can be re-written in a form linear in the Riemann tensor.
This allowed to derive a universal  form for the boundary term in a very general class of theories.

One of the interesting functionals of the curvature is the integrated conformal anomaly. The local form of the anomaly in four dimensions  was established in works of Duff and collaborators,  \cite{Capper:1974ic}. The general classification of the anomalies made in \cite{Deser:1993yx} considers two types of the anomaly. The anomaly of type A is
given by the Euler density while the anomaly of type B is constructed from the Weyl tensor $W_{\alpha\beta\mu\nu}$ and its covariant derivatives. In the presence of boundaries one may use the extrinsic curvature of the boundary to construct the conformal invariant quantities. More precisely, it is the traceless part $\hat{K}_{\mu\nu}=K_{\mu\nu}-\frac{1}{d-1}\gamma_{\mu\nu} K$
of the extrinsic curvature ($\gamma_{\mu\nu}$ is the induced metric on the boundary) that transforms homogeneously under conformal transformations, 
$\hat{K}_{\mu\nu}\rightarrow e^{\sigma}\hat{K}_{\mu\nu}$ if $g_{\mu\nu}\rightarrow e^\sigma g_{\mu\nu}$.
Thus, in $d$ dimensions the integrated conformal anomaly may have the following general form
\be
&&\int_{{\cal M}_d}\sqrt{g}\left\langle T_{\mu\nu}\right\rangle  g^{\mu\nu}=a~ \chi({{\cal M}_d})+b_k\int_{{\cal M}_d}\sqrt{\gamma} I_k(W)\nonumber \\
&&+b'_k\int_{\partial{\cal M}_d}\sqrt{\gamma}J_k(W, \hat{K})+
c_n \int_{\partial{\cal M}_d}\sqrt{\gamma} {\cal K}_n(\hat{K})\, ,
\lb{1}
\ee
where $\chi({{\cal M}_d})$ is the Euler number of manifold ${{\cal M}_d}$, $I_k(W)$ are conformal invariants constructed from the Weyl tensor, ${\cal K}_n(\hat{K})$ are polynomial of degree $(d-1)$ of the trace-free extrinsic curvature. In this note we suggest that in some appropriate normalization $b'_k=b_k$ and that the corresponding boundary term $J_k(W,\hat{K})$
is in fact the Hawking-Gibbons type term for the bulk action $I_k(W)$. In dimension $d=4$ this suggestion can be tested by comparing our result with the direct calculation 
performed recently by Fursaev \cite{Fursaev:2015wpa}  for free fields of various spin (for scalar fields this was done earlier by Dowker and Schofield \cite{Dowker:1989ue}).  We then extend our consideration to dimension $d=6$ and derive the exact form for the 
boundary terms $J_k(W,\hat{K})$. 

Thus, in the presence of boundaries the only new conformal charges which appear to emerge are $c_n$ that are  related to the conformal invariant expressions constructed from the 
trace free extrinsic curvature. The respective terms in the anomaly are interesting since they are present even in flat spacetime. However, as shows the example of scalar field in $d=4$ these charges may depend on the choice of the boundary conditions. Therefore, their invariant meaning is not very clear.
It would be interesting to associate these charges with certain structures which appear in the correlation functions of the CFT stress-energy tensor when  boundaries are present.
Answering this question, however, goes beyond the scope of the present short note.

\section{Gibbons-Hawking type boundary terms }
In this section we briefly review the construction given in \cite{Barvinsky:1995dp}  and then adapt it to the invariants constructed from the Weyl tensor.
This construction uses the fact that by introducing the auxiliary fields any function of the curvature can be re-written in the form which is linear in the Riemann tensor.
In the cases we are interested in this paper it is sufficient to add two auxiliary fields $U_{\alpha\beta\mu\nu}$ and $V_{\alpha\beta\mu\nu}$. The bulk terms then takes the form
\be
I_{bulk}=\int_{{\cal M}_d}\left(U^{\alpha\beta\mu\nu}R_{\alpha\beta\mu\nu}-U^{\alpha\beta\mu\nu}V_{\alpha\beta\mu\nu}+F(V)\right)\, ,
\lb{2}
\ee
where the exact form of $F(V)$ depends on the original form of the action. Then, according to  \cite{Barvinsky:1995dp} in order to cancel the normal derivatives of the metric variation on the boundary under variation of (\ref{2})  one should add the corresponding boundary term,
\be
I_{boundary}=-\int_{\partial{\cal M}_d}U^{\alpha\beta\mu\nu}P^{(0)}_{\alpha\beta\mu\nu}\, , \ \ P^{(0)}_{\alpha\beta\mu\nu}=n_\alpha n_\nu K_{\beta \mu}-n_\beta n_\nu K_{\alpha \mu}- n_\alpha n_\mu K_{\beta \nu}+ n_\beta n_\mu K_{\alpha \nu}\, .
\lb{3}
\ee
If the bulk invariant is expressed in terms of Weyl tensor only, the above procedure produces the following result for a manifold with boundary
\be
I[W]=\int_{{\cal M}_d}\left(U^{\alpha\beta\mu\nu}W_{\alpha\beta\mu\nu}-U^{\alpha\beta\mu\nu}V_{\alpha\beta\mu\nu}+F(V)\right)-\int_{\partial{\cal M}_d}U^{\alpha\beta\mu\nu}P_{\alpha\beta\mu\nu}\, ,
\lb{4}
\ee
where we introduced
\be
&&P_{\alpha\beta\mu\nu}=P^{(0)}_{\alpha\beta\mu\nu}-\frac{1}{d-2}(g_{\alpha\mu}P^{(0)}_{\beta\nu}-g_{\alpha\nu}P^{(0)}_{\beta\mu}-g_{\beta\mu}P^{(0)}_{\alpha\nu}+g_{\beta\nu}P^{(0)}_{\alpha\mu}) \lb{5} \\
&&+\frac{P^{(0)}}{(d-1)(d-2)}(g_{\alpha\mu}g_{\beta\nu}-g_{\alpha\nu}g_{\beta\nu})\, ,\nonumber \\
&&P^{(0)}_{\mu\nu}=n_\mu n^\alpha K_{\alpha\beta}+n_\mu n^\alpha K_{\alpha\nu}-K_{\mu\nu}-n_\mu n_\nu K\, , \ \ P^{(0)}=-2K\, ,\nonumber
\ee
where we used that $n^\alpha n^\beta K_{\alpha\beta}=0$.
$P_{\alpha\beta\mu\nu}$ has same symmetries as the Weyl tensor. In particular, $P^\alpha_{\ \mu \alpha\nu}=0$.

An interesting property of $P_{\alpha\beta\mu\nu}$ is that it does not change if we redefine extrinsic curvature,
\be
K_{\mu\nu}\rightarrow K_{\mu\nu}-\lambda \gamma_{\mu\nu}\, , \ \ P_{\alpha\beta\mu\nu} \rightarrow P_{\alpha\beta\mu\nu}\, ,
\lb{K}
\ee
 where $\gamma_{\mu\nu}=g_{\mu\nu}-n_\mu n_\nu$ is the induced metric on the boundary. Under the conformal transformations, $g_{\mu\nu}\rightarrow e^{2\sigma} g_{\mu\nu}$,
 the extrinsic curvature changes as
 $K_{\mu\nu}\rightarrow e^{\sigma}(K_{\mu\nu}-\gamma_{\mu\nu} n^\alpha\partial_\alpha\sigma)$. Therefore, the invariance (\ref{K}) indicates that
 $P_{\alpha\beta\mu\nu}$ is a conformal tensor  which transforms  homogeneously under the conformal rescaling of metric, $P_{\alpha\beta\mu\nu}\rightarrow e^{3\sigma} P_{\alpha\beta\mu\nu}$.
Invariance (\ref{K}) also means that $P_{\alpha\beta\mu\nu}$ can be rewritten entirely in terms of the trace free extrinsic curvature $\hat{K}_{\mu\nu}=K_{\mu\nu}-\frac{1}{d-1}\gamma_{\mu\nu}K$. The latter is of course consistent with the conformal symmetry of $P_{\alpha\beta\mu\nu}$.

\bigskip

Let us consider some example.

\medskip

\noindent{\bf 1. $I[W]=\int_{{\cal M}_d}\Tr (W^n)$. }
In this case we have 
\be
F(V)=\Tr (V^n)\, , \ \ V=W\, , \ \ U=n W^{n-1}\, .
\lb{6}
\ee
After resolving equations for $V$ and $U$ one finds for a manifold with boundary
\be
I[W]=\int_{{\cal M}_d}\Tr (W^n)-\int_{\partial{\cal M}_d}n\Tr(PW^{n-1})\, ,
\lb{6'}
\ee
where $P$ is defined in (\ref{5}).

\medskip

\noindent{\bf 2. $I[W]=\int_{{\cal M}_d} \Tr (W\Box W)$. }
In this case we have 
\be
F(V)=\Tr (V\Box V)\, , \ \ V=W\, , \ \ U=2\Box W\, 
\lb{7}
\ee
and after resolving equations for $V$ and $U$ we find
\be
I[W]=\int_{{\cal M}_d}\Tr (W\Box W)-2\int_{\partial{\cal M}_d}\Tr(P\Box W)\, .
\lb{7'}
\ee

\medskip

These examples will be useful in the subsequent sections.

\section{Conformal anomaly in $d=4$}
In four dimensions the local form of the anomaly is well-known
\be
&&\la T\ra=-\frac{a}{5760\pi^2}E_4+\frac{b}{1920\pi^2} \Tr W^2\, ,\nonumber \\
&&E_4=R_{\alpha\beta\mu\nu}R^{\alpha\beta\mu\nu}-4R_{\mu\nu}R^{\mu\nu}+R^2\, ,\nonumber \\
&&\Tr W^2=R_{\alpha\beta\mu\nu}R^{\alpha\beta\mu\nu}-2R_{\mu\nu}R^{\mu\nu}+\frac{1}{3}R^2\, ,
\lb{8}
\ee
In this normalization a scalar field has $a=b=1$. The integrated conformal anomaly contains the bulk integrals of the rhs of (\ref{8})
and some boundary terms. In particular, the bulk integral of $E_4$ is supplemented by some boundary terms  to form  a topological invariant, the Euler number,
\be
&&\chi[{{\cal M}_4}]=\frac{1}{32\pi^2}\int_{{\cal M}_4} E_4\nonumber \\
&&-\frac{1}{4\pi^2}\int_{\partial{\cal M}_4}(K^{\mu\nu}R_{n\mu n\nu}-K^{\mu\nu}R_{\mu\nu}-KR_{nn}+
\frac{1}{2}KR -\frac{1}{3}K^3+K\Tr K^2+\frac{2}{3}\Tr K^3)\, ,
\lb{9}
\ee
where $R_{\mu n\nu n}=R_{\mu\alpha\nu \beta }n^\alpha n^\beta$ and $R_{nn}=R_{\mu\nu}n^\mu n^\nu$. This form for the boundary terms was found in \cite{Dowker:1989ue}.

On the other hand, the integral of the Weyl tensor squared
should be supplemented by a boundary term as we explained in the previous section, see eq.(\ref{6'}) for $n=2$,
\be
\int_{{\cal M}_4}\Tr W^2-2\int_{\partial {\cal M}_4}\Tr (W P)\, .
\lb{10}
\ee
The properties of the Weyl tensor insure a simplification: 
$$\Tr (WP)=\Tr (W P^{(0)})=4W^{\mu\nu\alpha\beta}n_\mu n_\beta \hat{K}_{\nu\alpha}~. 
$$
As for the boundary term constructed from the traceless part of the  extrinsic curvature, in four dimensions there exists only one such term, $\Tr \hat{K}^3$.

Bringing everything together we arrive at the following form for the integrated conformal anomaly in four dimensions,
\be
\int_{{\cal M}_4}\la T\ra=-\frac{a}{180}\chi[{\cal M}_4]+\frac{b}{1920\pi^2}\left(\int_{{\cal M}_4} \Tr W^2-8\int_{\partial{\cal M}_4}W^{\mu\nu\alpha\beta}n_\mu n_\beta \hat{K}_{\nu\alpha}\right)+\frac{c}{280\pi^2}\int_{\partial{\cal M}_4} \Tr\hat{K}^3\, .
\lb{11}
\ee
The important point here is the balance between the bulk $\Tr W^2$ term and the boundary term $\Tr (WP^{(0)})$.
This  balance can be tested by comparing with the direct calculation performed recently by Fursaev \cite{Fursaev:2015wpa} for free fields of spin $s=0, \  1/2,\ 1$.
In all these cases an exact agreement with our general result (\ref{11}) is found. 

In the normalization used in (\ref{11}) the value of charge $c$, as
calculated in  \cite{Fursaev:2015wpa}, is
$c=1$ for scalar with Dirichlet boundary condition (and $c=7/9$ for a conformal Robin boundary condition), $c=5$ for the Dirac fermion and $c=8$ for a gauge vector field.

It should be noted that, in general, in the local form of the anomaly (\ref{8}) there may appear a total derivative term $\Box R$. If integrated it will produce a boundary term
$\int_{\partial{\cal M}_4}n^\mu \partial_\mu R$ which is not conformal invariant. No such term has appeared in the free field calculation of \cite{Fursaev:2015wpa}.
Apparently, the reason is that even though this total derivative term may be present in the local form of the anomaly, when integrated,   it is canceled by exactly same boundary term. It would be interesting to see whether the mechanism of this cancellation  is general.

\section{Conformal anomaly in $d=6$}
In a generic conformal field theory in $d=6$ the local form for the trace anomaly, modulo the total derivatives ($TD$),   is a combination of four different terms,
see \cite{Bastianelli:2000hi}, 
\be
\left\langle T\right\rangle={\cal A}=aE_6+b_1I_1+b_2I_2+b_3I_3+TD\, , 
\lb{12}
\ee
where $E_6$ is the Euler density in $d=6$ and we defined
\be
&&I_1=\Tr_1(W^3)=W_{\alpha\mu\nu\beta}W^{\mu\sigma\rho\nu}W_{\sigma\ \ \ \rho}^{\ \alpha\beta}\, ,\nonumber \\
&&I_2=\Tr_2(W^3)=W_{\alpha\beta}^{\ \ \mu\nu}W_{\mu\nu}^{\ \ \sigma\rho}W_{\sigma\rho}^{\ \ \alpha\beta}\, , \nonumber \\
&&I_3=\Tr (W\Box W)+\Tr_2(WXW)\, , \ \ X_{\alpha\beta}^{\ \ \mu\nu}=X^{[\mu}_{[\alpha}\delta^{\nu]}_{\beta]}\, , \ \ X^\mu_\nu=4R^\mu_\nu-\frac{6}{5}R\delta^\mu_\nu\, .
\lb{13}
\ee
In general there are two independent ways to define a product of Weyl tensors. That is why we have two different invariants $I_1$ and $I_2$.
Both invariants, formally,  can be written in the same  form  $\Tr W^3$ for each definition of the product. For the trace of a  product of two Weyl tensors
these two definitions are the same.

The list of possible total derivatives which may appear in the anomaly (\ref{12}) is available in \cite{Bastianelli:2000hi}. Whether all of them  contribute to the integrated conformal anomaly is an interesting question. We will, however, ignore them in our discussion.

In the integrated conformal anomaly the bulk integrals of  the rhs of (\ref{12}) are supplemented by certain boundary terms.
One group of these terms is such that in a combination with integral of $E_6$ it produces the topological Euler number of $6$-dimensional manifold with a boundary.
The exact form of these terms can be extracted from the general formula available in \cite{Dowker:1989ue}. More explicit expressions in higher dimensions (including  $d=6$) are available in a recent paper \cite{Herzog:2015ioa}.
The boundary terms that accompany
the bulk integrals of $I_k$, $k=1,\ 2, \ 3$ in the integrated conformal anomaly can be obtained by  same arguments as in section 2, see eqs.(\ref{6'}), (\ref{7'}).
The result for the integrated anomaly is\footnote{There may also exist an additional invariant, not included in (\ref{14}), that contains derivatives of extrinsic curvature, similarly to invariant $I_3$ in the bulk. It should be possibly to find exact expression for this invariant. We thank A. Waldron and K. Jensen for discussions on this point.} 
\be
&&\int_{{\cal M}_6}\la T\ra=a'\chi[{\cal M}_6]+b_1\left(\int_{{\cal M}_6} \Tr_1 W^3-3\int_{\partial{\cal M}_6}\Tr_1(PW^2)\right)\lb{14} \\
&&+b_2\left(\int_{{\cal M}_6} \Tr_2 W^3-3\int_{\partial{\cal M}_6}\Tr_2(PW^2)\right)\nonumber \\
&&+b_3\left(\int_{{\cal M}_6} \Tr (W\Box W)-  2\int_{\partial{\cal M}_6}\Tr(P\Box W)+\int_{{\cal M}_6} \Tr_2(WXW)- \int_{\partial{\cal M}_6}\Tr_2(WQW)\right)\nonumber \\
&&+c_1\int_{\partial{\cal M}_6}\Tr \hat{K}^2\Tr\hat{K}^3+c_2\int_{\partial{\cal M}_6}\Tr \hat{K}^5\, ,\nonumber
\ee
where $a'$ is the appropriately normalized charge $a$ and  we introduced
\be
Q_{\alpha\beta}^{\ \ \mu\nu}=Q^{[\mu}_{[\alpha}\delta^{\nu]}_{\beta]}\, , \ \ Q_{\mu\nu}=4P^{(0)}_{\mu\nu}-\frac{6}{5}P^{(0)}g_{\mu\nu}\, .
\lb{15}
\ee
It would be interesting to test (\ref{14}) for free fields  or in a calculation using holography.

\section{Conformal anomaly in odd dimensions}
In odd dimensions the local conformal anomaly vanishes since there are no local invariants constructed from the curvature with appropriate scaling.
However, the integrated conformal anomaly is, in general, non-vanishing due to the boundary terms. These terms can be constructed  from the intrinsic curvature of the boundary, the extrinsic curvature and the bulk curvature. Let us consider some examples.

\bigskip

\noindent {\bf Dimension $d=3$}. If the boundary is a two-dimensional compact surface   there are two possible  boundary terms which are conformal invariant: the Euler number of the boundary and the trace of a square of the trace-free extrinsic curvature, $\hat{K}_{ij}=K_{\ij}-\frac{1}{d-1}\gamma_{ij}K$, here we use the projection on the boundary so that indexes $i,j$ are along the two-dimensional surface.
Thus in this case the possible form of the anomaly is
\be
\int_{{\cal M}_3}\la T\ra =\frac{c_1}{96}\chi[\partial{\cal M}_3]+\frac{c_2}{256\pi}\int_{\partial{\cal M}_3}\Tr\hat{K}^2\, ,
\lb{16}
\ee
where $\chi[\partial{\cal M}_3]=\frac{1}{4\pi}\int_{\partial{\cal M}_3} \hat{R}$ is the Euler number of the boundary, $\hat{R}$ is the  scalar curvature of the boundary metric.
For a conformal scalar we find $c_1=-1$ and $c_2=1$ for the Dirichlet boundary condition and $c_1=1$ and $c_2=1$ for the conformal Robin condition, $(\partial_n+K/4)\phi|_{\partial{\cal M}_3}=0$. We used the form for the heat kernel coefficient $a_3$ found in \cite{Branson:1995cm} when derived this result for the scalar field.

\bigskip

\noindent {\bf Dimension $d=5$}. In five dimensions there are quite a few possible conformal invariants. We list some of them below
\be
&&\int_{{\cal M}_5}\la T\ra =c_1\chi[\partial{\cal M}_5]+\int_{\partial{\cal M}_5}[c_2 \Tr W^2 +c_3 W_{\alpha n\beta n}W^{\alpha \ \beta }_{\ n \ n}+c_4 W^{\alpha\mu\beta\nu}\hat{K}_{\alpha\beta}\hat{K}_{\mu\nu}\nonumber \\
&&+c_5 (\Tr \hat{K}^2)^2+
c_6\Tr \hat{K}^4+c_7\Tr(\hat{K}{\cal D}\hat{K})+c_8 W_{n\alpha\beta\mu}W_n^{\  \alpha\beta\mu} ]\, ,
\lb{17}
\ee
where $W_{\alpha n \beta n}=W_{\alpha\mu \beta \nu} n^\mu n^\nu$ and 
\be
&&{\cal D}\hat{K}_{ij}=\hat{\nabla}^2 \hat{K}_{ij}+\frac{1}{6}\hat{R}\hat{K}_{ij}-(\hat{R}_{ik}\hat{K}^k_j+\hat{R}_{jk}\hat{K}^k_i)+\frac{1}{2}\gamma_{ij}\hat{R}^{km}\hat{K}_{km}\nonumber \\
&&-\frac{2}{3}(\hat{\nabla}_i\hat{\nabla}_k \hat{K}^k_j+ \hat{\nabla}_j\hat{\nabla}_k \hat{K}^k_i)+\frac{1}{3}\gamma_{ij}\hat{\nabla}^k\hat{\nabla}^m\hat{K}_{km}
\lb{18}
\ee
is the conformal operator acting on trace-free symmetric tensors in four dimensions, see for instance  \cite{Achour:2013afa}, the covariant derivatives are defined with respect to intrinsic metric on the boundary. We can not exclude that some other conformal invariants constructed from the derivatives of extrinsic curvature may exist\footnote{I thank C. Berthiere for discussions on this point.}.
It would be certainly interesting to compute the boundary charges $c_k$ in some particular cases in five dimensions.

\section{Conclusions}

In this note we suggested that the boundary terms that correspond to the anomaly of type B and appear  in the integrated conformal anomaly are of the Gibbons-Hawking type.
In four dimensions we have found the corresponding term to agree with a free field calculation in \cite{Fursaev:2015wpa}. In six dimensions the boundary terms corresponding to the 
conformal invariants $I_1$, $I_2$ and $I_3$ have been found explicitly. 

On the other hand, the local anomaly of type A, if integrated,  is completed by a boundary term to reproduce the topological Euler number for a manifold with boundaries. A general formula for   the boundary term in  the Euler topological characteristic   can be  found in \cite{Dowker:1989ue}  and \cite{Herzog:2015ioa}.

Taking the importance of the problem it would be very interesting to formulate a classification theorem for  the conformal boundary invariants similar to the one
that exists for the bulk conformal invariants. For a recent progress in this direction see \cite{Glaros:2015pfa}, where a number of conformal hypersurface invariants
were proposed including the extrinsic analogs of
the Paneitz operator and all higher Laplacian powers. It would be nice to make a bridge between these mathematical studies and the physics approach developed in this note.

On the physics side, an interesting open question which seems to remain is what is the invariant meaning of the boundary charges $c_k$ that correspond to the boundary terms constructed from the trace free part of the extrinsic curvature. These boundary terms do not  have a bulk counterpart and the corresponding charges appear to be unrelated to any conformal charges in the bulk.
A reasonable  guess is that $c_k$ are some new charges which characterize the conformal theory in the presence of a boundary. Possibly, they are related to certain structures in the
correlation functions of the CFT stress-energy tensor when they are restricted to the boundary. 
A similar question arises for the boundary charges in odd dimensions.
It would be interesting to understand better these  issues.

\section*{Acknowledgements} 
I thank D. Fursaev for communications regarding his calculation and A. Waldron and K. Jensen for comments.


\newpage
\begin{thebibliography}{999}

{\frenchspacing \parskip=2mm
 
\bibitem{Gibbons:1976ue} 
  G.~W.~Gibbons and S.~W.~Hawking,
  Phys.\ Rev.\ D {\bf 15}, 2752 (1977).
  
\bibitem{Barvinsky:1995dp} 
  A.~D.~Barvinsky and S.~N.~Solodukhin,
  Nucl.\ Phys.\ B {\bf 479}, 305 (1996)
  [gr-qc/9512047].
  
 
\bibitem{Capper:1974ic} 
  D.~M.~Capper and M.~J.~Duff,
  Nuovo Cim.\ A {\bf 23}, 173 (1974).\\
   S.~Deser, M.~J.~Duff and C.~J.~Isham,
  Nucl.\ Phys.\ B {\bf 111}, 45 (1976).\\
  M.~J.~Duff,
  Class.\ Quant.\ Grav.\  {\bf 11}, 1387 (1994)
  [hep-th/9308075].


\bibitem{Deser:1993yx} 
  S.~Deser and A.~Schwimmer,
  Phys.\ Lett.\ B {\bf 309}, 279 (1993)
  [hep-th/9302047].
 
 
  
\bibitem{Fursaev:2015wpa} 
  D.~Fursaev,
  ``Conformal anomalies of CFT's with boundaries,''
  arXiv:1510.01427 [hep-th].
  
\bibitem{Dowker:1989ue} 
  J.~S.~Dowker and J.~P.~Schofield,
  J.\ Math.\ Phys.\  {\bf 31}, 808 (1990).
 
\bibitem{Bastianelli:2000hi}
  F.~Bastianelli, S.~Frolov and A.~A.~Tseytlin,
  JHEP {\bf 0002} (2000) 013
  [hep-th/0001041].
  
\bibitem{Herzog:2015ioa} 
  C.~P.~Herzog, K.~W.~Huang and K.~Jensen,
  ``Universal Entanglement and Boundary Geometry in Conformal Field Theory,''
  arXiv:1510.00021 [hep-th].
  
\bibitem{Glaros:2015pfa} 
  M.~Glaros, A.~R.~Gover, M.~Halbasch and A.~Waldron,
  ``Singular Yamabe Problem Willmore Energies,''
  arXiv:1508.01838 [math.DG].\\
  A.~R.~Gover and A.~Waldron,
  ``Conformal hypersurface geometry via a boundary Loewner-Nirenberg-Yamabe problem,''
  arXiv:1506.02723 [math.DG].
  
  
  


\bibitem{Branson:1995cm} 
  T.~P.~Branson, P.~B.~Gilkey and D.~V.~Vassilevich,
  Boll.\ Union.\ Mat.\ Ital.\  {\bf 11B}, 39 (1997)
  [hep-th/9504029].

\bibitem{Achour:2013afa} 
  J.~Ben Achour, E.~Huguet and J.~Renaud,
  Phys.\ Rev.\ D {\bf 89}, 064041 (2014)
  [arXiv:1311.3124 [gr-qc]].
 
}


\end{thebibliography}
\end{document}